\documentclass[11pt, prd, aps, nofootinbib, superscriptaddress,
tightenlines]{revtex4}

\usepackage{amssymb}
\usepackage{amsmath}
\usepackage{color}
\usepackage{latexsym}
\usepackage{theorem}
\usepackage{graphicx}
\usepackage{subfigure}
\usepackage{graphics}
\usepackage{epsfig}
\usepackage{makeidx}
\usepackage{showidx}

\newbox\grsign \setbox\grsign=\hbox{$>$} \newdimen\grdimen \grdimen=\ht\grsign
\newbox\simlessbox \newbox\simgreatbox
\setbox\simgreatbox=\hbox{\raise.5ex\hbox{$>$}\llap
     {\lower.5ex\hbox{$\sim$}}}\ht1=\grdimen\dp1=0pt
\setbox\simlessbox=\hbox{\raise.5ex\hbox{$<$}\llap
     {\lower.5ex\hbox{$\sim$}}}\ht2=\grdimen\dp2=0pt
\def\simgreat{\mathrel{\copy\simgreatbox}}
\def\simless{\mathrel{\copy\simlessbox}}
\def\mmsun{\mathrm{M}_\odot}

\newcommand{\codename}[1]{\texttt{#1}}

\begin{document}

\title{Cactus Framework: Black Holes to Gamma Ray Bursts}

\author{Erik Schnetter}
\affiliation{Center for Computation \& Technology, 216 Johnston Hall,
  Louisiana State University, LA 70803, USA}
\homepage{http://www.cct.lsu.edu/}
\affiliation{Department of Physics and Astronomy, 202 Nicholson Hall,
  Louisiana State University, Baton Rouge, LA 70803, USA}

\author{Christian D. Ott}
\affiliation{Steward Observatory and Department of Astronomy, The
  University of Arizona, 933 N. Cherry Ave., Tucson, AZ 85721, USA}

\author{Gabrielle Allen}
\affiliation{Center for Computation \& Technology, 216 Johnston Hall,
  Louisiana State University, LA 70803, USA}

\author{Peter Diener}
\affiliation{Center for Computation \& Technology, 216 Johnston Hall,
  Louisiana State University, LA 70803, USA}
\affiliation{Department of Physics and Astronomy, 202 Nicholson Hall,
  Louisiana State University, Baton Rouge, LA 70803, USA}

\author{Tom Goodale}
\affiliation{Center for Computation \& Technology, 216 Johnston Hall,
  Louisiana State University, LA 70803, USA}
\affiliation{School of Computer Science, Cardiff University, The Parade,
  Cardiff, CF24 3AA, UK}

\author{Thomas Radke}
\affiliation{Max-Planck-Institut f\a"ur Gravitationsphysik,
  Albert-Einstein-Institut, Am M\a"uhlenberg 1, D-14476 Golm, Germany}

\author{Edward Seidel}
\affiliation{Center for Computation \& Technology, 216 Johnston Hall,
  Louisiana State University, LA 70803, USA}

\author{John Shalf}
\affiliation{Lawrence Berkeley National Laboratory, 1 Cyclotron Rd.,
  Berkeley, CA 94720, USA}

\date{June 30, 2007}

\begin{abstract}
  Gamma Ray Bursts (GRBs) are intense narrowly-beamed flashes of
  $\gamma$-rays of cosmological origin.  They are among the most
  scientifically interesting astrophysical systems, and the riddle
  concerning their central engines and emission mechanisms is one of
  the most complex and challenging problems of astrophysics today.
  In this article we outline our petascale approach to the GRB problem
  and discuss the computational toolkits and numerical codes that are
  currently in use and that will be scaled up to run on emerging
  petaflop scale computing platforms in the near future.
  
  Petascale computing will require additional ingredients over
  conventional parallelism.  We consider some of the challenges which
  will be caused by future petascale architectures, and discuss our
  plans for the future development of the Cactus framework and its
  applications to meet these challenges in order to profit from these
  new architectures.
\end{abstract}

\maketitle

\section{Current challenges in relativistic astrophysics 
and the Gamma-Ray Burst problem}

Ninety years after Einstein first proposed his General
Theory of Relativity (GR), astrophysicists are more 
than ever and in greater detail probing into regions
of the universe where gravity is very strong and where,
according to GR's geometric description, the curvature
of spacetime is large. 

The realm of strong curvature is notoriously difficult to 
investigate with conventional observational astronomy,
and some phenomena might bear no observable electro-magnetic
signature at all and may only be visible in neutrinos (if sufficiently
close to Earth) or in gravitational
waves --- ripples of spacetime itself which are predicted by
Einstein's GR\@. Gravitational waves have not been
observed directly to date, but gravitational-wave 
detectors (e.g., LIGO~\cite{ligoweb}, 
GEO~\cite{geoweb}, VIRGO~\cite{virgoweb}) are in
the process of reaching sensitivities sufficiently
high to observe interesting astrophysical phenomena.

\begin{figure}
  \includegraphics[width=0.445\textwidth]{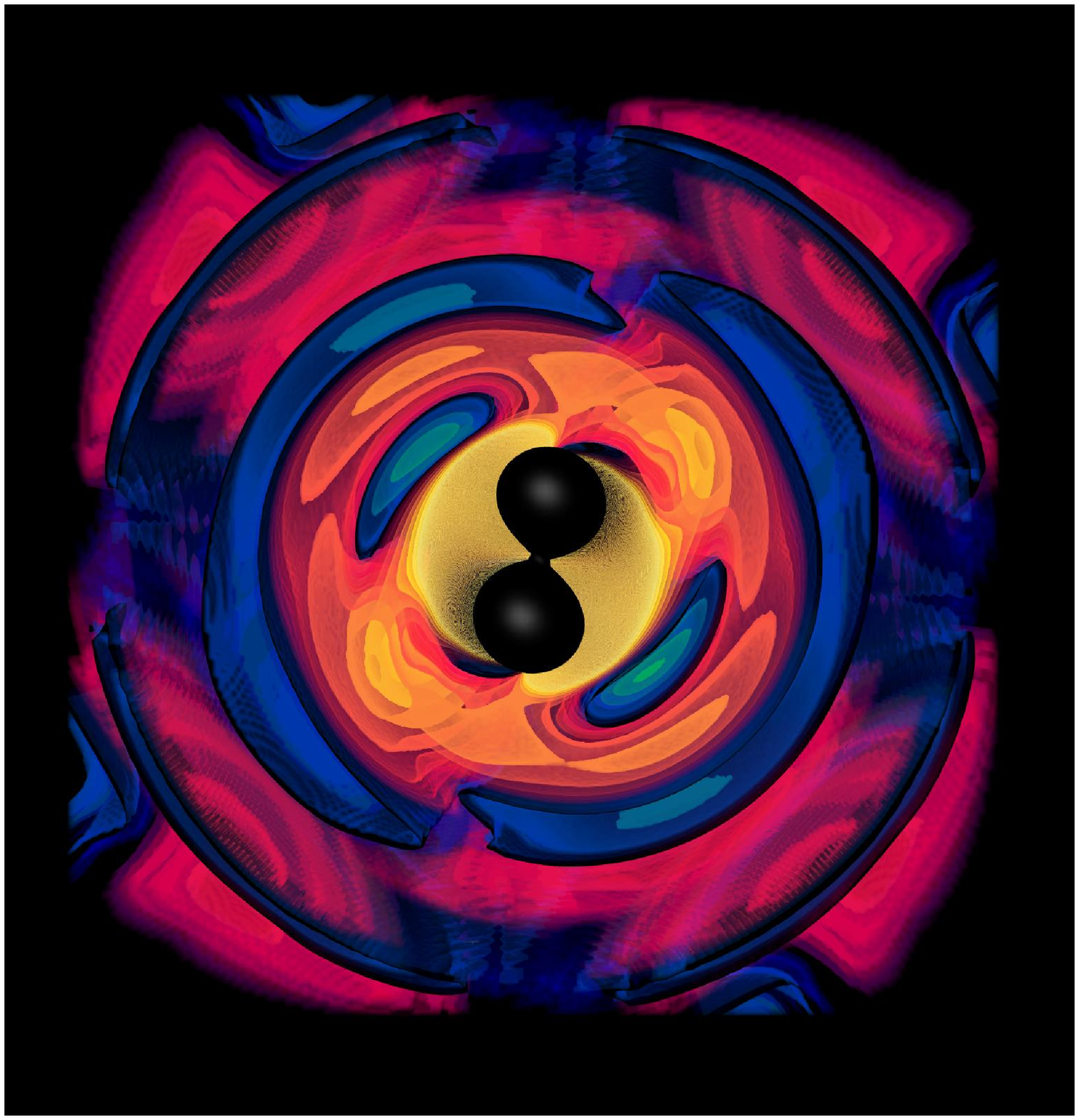}
  \includegraphics[width=0.545\textwidth]{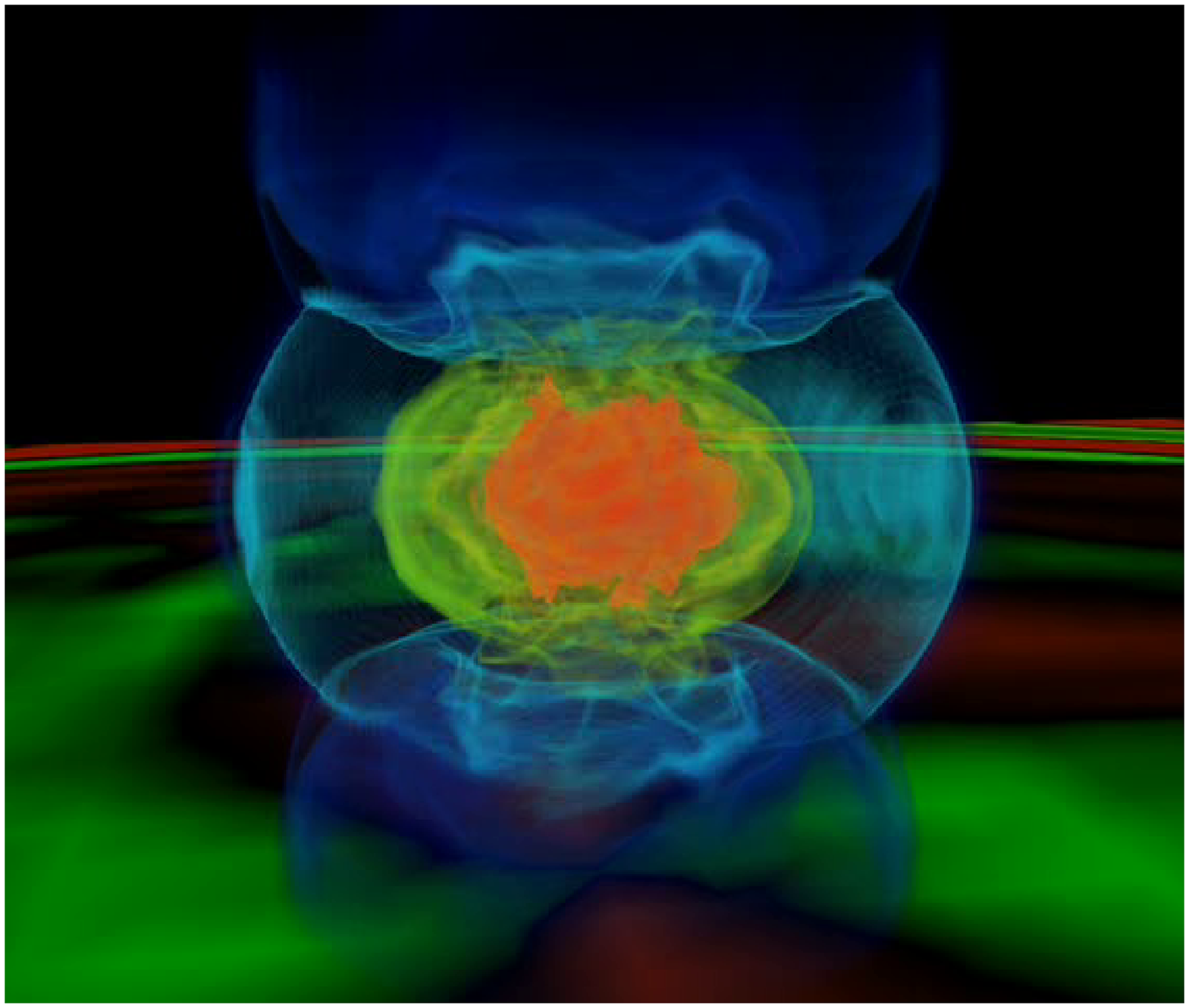}
  \caption{\emph{Left:} Gravitational waves and horizons in a binary
    black hole inspiral simulation.  Simulation by AEI/CCT
    collaboration, image by W. Benger (CCT/AEI/ZIB).  \emph{Right:} 
    Rotationally deformed protoneutron star formed in the iron core
    collapse of an evolved massive star. Shown are a volume rendering
    of the rest-mass density and a 2D rendition of 
    outgoing gravitational waves. Simulation by \cite{Ott07prl},
    image by R. K\"ahler.}
  \label{fig:illustrative}
\end{figure}

Until gravitational-wave astronomy becomes reality, 
astrophysicists must rely on computationally and conceptually 
challenging large-scale numerical simulations in order 
to grasp the details of the energetic processes
occurring in regions of strong curvature that
are shrouded from direct observation in the electromagnetic
spectrum by intervening matter or that have 
little or no electromagnetic signature at all.
Such astrophysical systems and phenomena include the 
birth of neutron stars (NSs) or
black holes (BHs) in collapsing evolved massive stars,
coalescence of compact\footnote{The term
``compact'' refers to the compact-stellar nature
of the binary members in such systems: white dwarfs,
neutron stars, black holes.} binary systems, 
gamma-ray bursts (GRBs), active galactic nuclei 
harboring supermassive black holes, pulsars, and
quasi-periodically oscillating NSs (QPOs). In figure~\ref{fig:illustrative}
we present example visualizations of binary BH and
stellar collapse calculations carried out by our groups.

From these,
GRBs, intense narrowly-beamed flashes of $\gamma$-rays
of cosmological origin, are among the most scientifically interesting
and the riddle concerning their central engines and
emission mechanisms is one of the most complex
and challenging problems of astrophysics today.

GRBs last between 0.5--1000~secs, with a bimodal distribution of 
durations~\cite{meszaros:06}, indicating two distinct classes
of mechanisms and central engines. The short-hard (duration $\simless 2$~secs)
group of GRBs (hard, because their $\gamma$-ray spectra peak at shorter wavelength) 
predominantly occurs in elliptical galaxies 
with old stellar populations at moderate astronomical 
distances~\cite{wb:06,meszaros:06}. The energy released in 
a short-hard GRB and its duration suggest~\cite{wb:06,meszaros:06} 
a black hole with a $\sim$0.1 solar-mass ($\mmsun$) accretion disk 
as the central engine. Such a BH--accretion-disk system 
is likely to be formed by the coalescence of NS--NS or 
NS--BH systems~(e.g., \cite{setiwan:06}).

Long-soft (duration $\sim$2--1000~secs) 
GRBs on the other hand seem to occur exclusively in
the star-forming regions of spiral or irregular 
galaxies with young stellar populations and low 
metallicity\footnote{The metallicity of an astrophysical 
object is the mass fraction 
in chemical elements  other than hydrogen and helium. 
In big-bang nucleosynthesis
only hydrogen and helium were formed. All other elements are
ashes of nuclear burning processes in stars.} 
Observations that have recently become available 
(see \cite{meszaros:06} for reviews) 
indicate features in the x-ray and optical
afterglow spectra and luminosity evolutions 
of long-soft GRBs that show similarities with
spectra and light curves obtained from Type-Ib/c 
core-collapse supernovae whose progenitors are
evolved massive stars ($M \simgreat 25\, \mmsun$) that
have lost their extended hydrogen envelopes and probably also 
a fair fraction of their helium shell. These observations
support the collapsar model~\cite{wb:06} 
of long-soft GRBs that envisions a stellar-mass
black hole formed in the aftermath of a stellar
core-collapse event with a massive $\sim$1~$\mmsun$ 
rotationally-supported accretion disk as the central engine,
powering the GRB jet that punches through the compact
and rotationally-evacuated polar stellar envelope reaching
ultra-relativistic velocities~\cite{meszaros:06}.

Although  observations are
aiding our theoretical understanding, much that is said about
the GRB central engine will remain speculation until it is 
possible to  self-consistently  
model (i) the processes that lead to the formation of 
the GRB central engine and
(ii) the way the central engine utilizes gravitational 
(accretion) and rotational energy to launch the GRB jet
via magnetic stresses and/or polar neutrino pair-annihilation
processes. The physics necessary in such a model
includes general relativity, relativistic magneto-hydrodynamics, 
nuclear physics (describing nuclear reactions and the 
equation of state of dense matter), neutrino physics 
(weak interactions), and neutrino and photon radiation transport.
In addition, it is necessary to adequately resolve physical
processes with characteristic scales from $\sim$100~meters
near the central engine to $\sim$5--10~million~kilometers, the
approximate radius of the collapsar progenitor star.

\subsection{GRBs and petascale computing}

The complexity of the GRB central engine and its
environs requires a multi-physics, multi-length-scale
approach that cannot be fully realized on 
present-day computers.  Computing at multiple sustained 
petaflops of performance will allow us
to tackle the full GRB problem and provide complete 
numerical models whose output can be compared with 
observations.

In this article we outline our petascale approach
to the GRB problem and discuss the computational
toolkits and numerical codes that are currently
in use and that will be scaled up to run on emerging 
petaflop scale computing platforms in the
near future. 

Any comprehensive approach to GRBs must naturally draw 
techniques and tools from both numerical relativity 
and core-collapse supernova and neutron star theory. 
Hence, much of the work presented and suggested here builds
upon the dramatic progress that has been made
in these fields in the past decade.
In numerical relativity, immense improvements in the long-term 
stability of 3D GR vacuum and hydrodynamics evolutions
(e.g., \cite{Alcubierre02a,Pretorius:2005gq})
allow for the first time long-term stable binary black 
hole merger, binary neutron star merger, neutron star
and evolved massive star collapse calculations.
Supernova theory, on the other hand, has made giant
leaps from spherically symmetric (1D) models with 
approximate neutrino radiation transport in the 
early 1990s, to Newtonian or approximate-GR
to 2D and the first 3D~\cite{fryerwarren:04} 
calculations, including detailed neutrino and 
nuclear physics and energy-dependent multi-species 
Boltzmann neutrino transport~\cite{buras:06b} or 
neutrino flux-limited diffusion~\cite{burrows:07a}
and magneto-hydrodynamics~\cite{burrows:07b}.

As we shall discuss, our present suite of terascale codes,
comprised of the spacetime evolution code \codename{Ccatie}
and the GR hydrodynamics code \codename{Whisky}
can be and has already been applied to the realistic modeling
of the inspiral and merger phase of NS--NS and NS-BH binaries, 
to the collapse of polytropic (cold) supermassive NSs, and 
to the collapse and early post-bounce phase of a core-collapse 
supernova or a collapsar. As the codes will be upgraded and
readied for petascale, the remaining physics modules
will be developed and integrated. In particular, 
energy-dependent neutrino transport and magneto-hydrodynamics,
both likely to be crucial for the GRB central engine, will
be given high priority.

To estimate roughly the petaflopage and petabyteage 
required for a full collapsar-type GRB calculation, we assume a 
Berger-Oliger-type~\cite{Berger84} adaptive-mesh
refinement setup with 16 refinement levels, resolving features
with resolution of 10,000~km down to 100~m across a domain
of 5 million cubic km. To simplify things, we assume that each 
level of refinement has twice the resolution as the previous
level and covers approximately half the domain. Taking 
a base grid size of 1024$^3$ and 512 3D grid functions,
storing the curvature, and radiation-hydrodynamics data
on each level, we estimate a total memory consumption
of $\sim$0.0625 PB (64 TB). To obtain an estimate of 
the required sustained petaflopage, we first compute
the number of time steps that are necessary to evolve
for 100~s in physical time. Assuming a time step that
is half the light-crossing time of each grid cell on each
individual level, the base grid has to be evolved for
$\sim$6000 time steps, while the finest grid will have to
be evolved for $2^{16-1} \times 6000$ individual time steps.
\codename{Ccatie} plus \codename{Whisky} require approximately
10k flops grid point per time step. When we assume that
additional physics (neutrino and photon radiation transport, 
magnetic fields; some of which may be evolved with different and varying 
time step sizes) requires on average additional 22k flops, 
one time step of one refinement level requires 
$10^{-5}$~petaflops. Summing up over all levels and
time steps, we arrive at a total petaflopage of
$\sim$13 million. On a machine with 2 petaflops sustained,
the runtime of the simulation would come to $\sim$75~days.
GRBs pose a true petascale problem.

\section{The Cactus framework}
\label{sec:cactus}

To reduce the development time for creating simulation codes and
encourage code reuse, researchers have created computational
frameworks such as the Cactus Framework~\cite{Goodale02a, cactusweb1}.
Such modular component-based frameworks allow
scientists and engineers to develop their own application modules 
(CS services or physics solvers) and
assemble them with a body of existing code components to create applications that
solve complex multiphysics computational problems. 
Cactus provides tools ranging from basic computational
building blocks to complete toolkits that can be used to solve a range
of application problems.  Cactus runs on a wide range of hardware
ranging from desktop PC's, large supercomputers, to 'grid'
environments.  The Cactus Framework and core toolkits are distributed
with an open source license from the Cactus
website~\cite{cactusweb1}, are fully documented, and are maintained
by active developer and user communities.

The Cactus Framework consists
of a central infrastructure (``flesh'') and components (``thorns'').  The
flesh has minimal computational functionality and serves as a
module manager, coordinating the flow of data between  the different components to
perform specific tasks.  The components or ``thorns'' perform tasks
ranging from setting up a computational grid, decomposing the
grid for parallel processing, setting up coordinate
systems, boundary and initial conditions, communication of data from
one processor to another, solving partial differential equations, to
input and output and streaming of visualization data. One standard set of thorns 
is distributed as the Cactus Computational Toolkit to provide basic 
functionality for computational science.

Cactus was originally designed for scientists and engineers to
collaboratively develop large-scale,
parallel scientific codes which would be run on laptops and workstations (for development)
and large supercomputers (for production runs). 
The Cactus thorns are organized in a manner that provides a clear separation
of the roles and responsibilities between the ``expert computer scientists'' who implement
complex parallel abstractions (typically in C or C++), and the ``expert mathematicians
and physicists'' who program thorns that look like serial blocks 
of code (typically, in F77, F90, or C) implementing complex numerical algorithms. 
Cactus provides the 
basic parallel framework supporting several different codes in the numerical relativity community
used for modeling black holes, neutron and boson stars and gravitational waves. This has lead to 
over 150 scientific publications in numerical relativity which have used Cactus and the establishment
of a Cactus Einstein Toolkit of shared community thorns. Other fields of science and engineering 
are also using the Cactus Framework, including Quantum Gravity, Computation Fluid Dynamics, 
Computational Biology, Coastal Modeling, Applied Mathematics etc, and in some of these 
areas community toolkits and shared domain specific tools and interfaces are emerging. 

Cactus provides a range of advanced development and runtime tools including; 
an HTTPD thorn that incorporates a simplified web server into the simulation allowing for
real-time monitoring and steering through any web interface; a general timer infrastructure
for users to easily profile and optimize their codes; visualization readers and writers for
scientific visualization packages; and interfaces to Grid Application Toolkits for developing 
new scientific scenarios taking advantage of distributed computing resources.

Cactus is highly portable.  Its build system detects
differences in machine architecture and compiler features,
using automatic detection where possible and a database of known
system environments where auto-detection is impractical --- e.g.\ which
libraries are necessary to link Fortran and C code together for a particular system architecture
or facility.  Cactus
runs on nearly all variants of the Unix operating system,
Windows platform, and a number of microkernels including Catamount and BlueGene CNK.
Codes written using Cactus have been run on some of the fastest
computers in the world, such as the Japanese Earth Simulator and the
IBM BlueGene/L, Cray X1e, and the Cray XT3/XT4 
systems~\cite{OLI_IPDPS07,OLI03,OLI04_SC,CART06,KAMLIISWC05,HFAST_SC2005}
Cactus' flexible and robust build system enables developers 
to write and test code on their laptop computers,
and then deploy the very same code on the full scale systems with very little effort.

\section{Spacetime and hydrodynamics codes}

\subsection{\protect\codename{Ccatie}: Spacetime evolution}
\label{sec:ccatie}

In strong gravitational fields, such as in the presence of neutron
stars or black holes, it is necessary to solve the full Einstein
equations.  Our code employs a $3+1$ decomposition~\cite{Arnowitt62,York79}, 
which renders the four-dimensional spacetime equations into
hyperbolic time-evolution equations in three dimensions, plus a set of
constraint equations which have to be satisfied by the initial
condition.  The equations are discretized using high order finite
differences with adaptive mesh refinement and using Runge--Kutta time
integrators, as described below.

The time evolution equations are formulated using a variant of the BSSN
formulation described in \cite{Alcubierre99d} and coordinate conditions
described in \cite{Alcubierre02a} and \cite{Baker:2006mp}.
These are a set of $25$ coupled partial differential equations which
are first order in time and second order in space.  One central
variable describing the geometry is the three-metric $\gamma_{ij}$,
which is a symmetric positive definite tensor defined everywhere in
space, defining a scalar product which defines distances and angles.

\codename{Ccatie} contains the formulation and discretization of
the right hand sides of the time evolution equations.  Initial data
and many analysis tools, as well as time integration and
parallelisation, are handled by other thorns.  The current state of
the time evolution, i.e., the three-metric $\gamma_{ij}$ and related
variables, are communicated into and out of \codename{Ccatie} using a
standard set of variables (which is different from \codename{Ccatie}'s
evolved variables), which makes it possible to combine unrelated
initial data solvers and analysis tools with \codename{Ccatie}, or to
replace \codename{Ccatie} by other evolution methods, while reusing
all other thorns.




A variety of initial conditions are provided by the \codename{Ccatie} 
thorn, ranging from simple
test cases, analytically known and perturbative solutions
to binary systems containing neutron stars and black holes.

The numerical kernel of \codename{Ccatie} has been hand-coded and
extensively optimised for FPU performance where the greatest part of 
the computation lies.  (Some analysis methods can be
similarly expensive and have been similarly optimised, e.g.\ the
calculation of gravitational wave quantities.)  The Cactus/Ccatie
combination has won various awards for performance.\footnote{See
  \url{http://www.cactuscode.org/About/Prizes}}

\subsection{\codename{Whisky}: General relativistic hydrodynamics}
\label{sec:whisky}

The \codename{Whisky} code \cite{Baiotti04, whiskyweb} is a
GR hydrodynamics code  originally developed 
under the auspices of the European Union research training
network ``Sources of Gravitational Waves''~\cite{eunetworkweb}.

While \codename{Ccatie} in combination with \codename{Cactus}'s
time-integration methods provides the time evolution of the
curvature part of spacetime, \codename{Whisky} evolves the
``right-hand side'', the matter part, of the Einstein equations.
The coupling of curvature with matter is handled by \codename{Cactus}
via a transparent and generic interface, providing for
modularity and interchangeability of curvature and matter
evolution methods. \codename{Whisky} is also fully integrated
with the \codename{Carpet} mesh refinement driver discussed
in \S\ref{section:carpet}.

\codename{Whisky} implements the equations of GR hydrodynamics
in a semi-discrete fashion,
discretising only in space and leaving the explicit 
time integration to \codename{Cactus}. The update terms
for the hydrodynamic variables are computed via flux-conservative 
finite-volume methods exploiting the characteristic structure 
of the equations of GR hydrodynamics. Multiple dimensions are handled via
directional splitting. 
Fluxes are computed via piecewise-parabolic cell-interface 
reconstruction and approximate Riemann solvers to provide
right-hand side data that are accurate to (locally) third-order 
in space and first-order in time. High temporal accuracy
is obtained via Runge-Kutta-type time-integration cycles handled
by \codename{Cactus}.

\codename{Whisky} has found extensive use in the study of 
neutron star collapse to a black hole,
incorporates matter excision techniques for stable
non-vacuum BH evolutions, has
been applied to BH -- NS systems,
and NS rotational instabilities.

In a recent upgrade, \codename{Whisky} has been endowed with the
capability to handle realistic finite-temperature nuclear equations
of state and to approximate the effects of electron capture
on free protons and heavy nuclei in the collapse phase of core-collapse
supernovae and/or collapsars~\cite{Ott07prl}. In addition,
a magneto-hydrodynamics version of \codename{Whisky}, \codename{WhiskyMHD},
is approaching the production stage~\cite{Giacomazzo:2007ti}.

\section{Parallel implementation and mesh refinement}

In Cactus, infrastructure components providing storage handling,
parallelisation, mesh refinement, and I/O methods are implemented by
thorns in the same manner that numerical components provide boundary
conditions or physics components provide initial data.  The component
which defines the order in which the time evolution is orchestrated, and
implements the parallel execution model is called the \emph{driver}.

Cactus has generally been used to date for calculations based upon explicit
finite difference methods.  Each simulation routine is invoked with a
block of data --- e.g.\ in a 3-dimensional simulation the routine is
passed a cuboid, in a 2-dimensional simulation a rectangle --- and
integrates the data in this block forward in time.  In general, the simulation
routines only operate on the data they are handed (no side-effects), do not
carry any internal state between invocations (statelessness), and do not
directly control the order of execution (they merely define constraints on orderings).
This organizes the Fortran and C (declarative) modules in a manner that is very
similar to functional/imperative constraints.  The imperative scheduling model
provides the Cactus \emph{drivers} considerable flexibility for the scheduling parallel execution.
Consequently, Cactus supports a number of execution paradigms (offered by 
different  \emph{driver}) without requiring any changes to the physics modules to 
accommodate them.

For example, in a single
processor simulation the block would consist of all the data for the
whole of the simulation domain, and in a multi-processor
simulation the domain is decomposed into smaller subdomains, where each
processor computes the block of data from its subdomain.  In a finite
difference calculation, the main form of communication between these
subdomains is on the boundaries, and this is done by \emph{ghost-zone
  exchange} whereby each subdomain's data-block is enlarged by the
nearest boundary data from neighbouring blocks.
The data from these ghost-zones is then exchanged
once per iteration of the simulation. Cactus can also support shared-memory, 
out-of-core, and AMR execution paradigms through different drivers while
still using the very same physics components.

There are several drivers available for Cactus.  In this paper we
present results using the unigrid \codename{PUGH}
driver and using the adaptive mesh refinement
(AMR) driver \codename{Carpet}.

\subsection{PUGH}

\begin{figure}
  \includegraphics[width=0.48\textwidth]{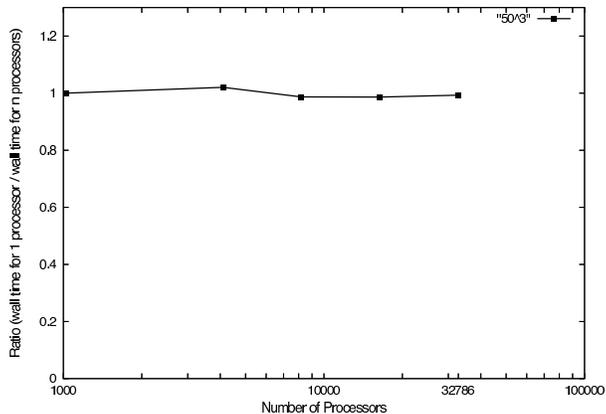}
  \caption{PUGH scaling result from BlueGene/L.  (See also figure 6 in
    the chapter ``Performance Characteristics of Potential Petascale
    Scientific Applications'' by Oliker et al.\ for similar
    Cactus/PUGH benchmarks.)}
  \label{fig:pugh-scaling}
\end{figure}

The Parallel UniGrid Hierarchy (\codename{PUGH}) driver was the first parallel
driver available in
Cactus.  PUGH decomposes the problem domain into one block per processor
using MPI for the ghost-zone exchange described above.  PUGH
has been successfully used on many architectures and has proven
scaling up to many thousands of processors~\cite{OLI_IPDPS07}.
Figure \ref{fig:pugh-scaling} shows benchmark results from
BlueGene/L.  (See also figure 6 in the chapter ``Performance
Characteristics of Potential Petascale Scientific Applications'' by
Oliker et al.)

\subsection{Adaptive Mesh Refinement with \codename{Carpet}}
\label{section:carpet}

\codename{Carpet}
\cite{Schnetter-etal-03b, carpetweb} is a driver which implements
Berger--Oliger mesh refinement~\cite{Berger84}.  Carpet refines parts
of the simulation domain in space and/or time by factors of two.  Each
refined region is block-structured, which allows for efficient
representations e.g.\ as Fortran arrays.

In addition to mesh refinement, Carpet also provides parallelism by
distributing grid functions onto processors, corresponding I/O methods
for ASCII and HDF5 output and for checkpointing and restart,
interpolation of values to arbitrary points, and reduction operations
such as norms and maxima.

Fine grid boundary conditions require interpolation in space.  This is
currently implemented up to seventh order, and fifth order is commonly
used.  When refining in time, finer grids take smaller time steps to
satisfy the local CFL criterion.  In this case, Carpet may need to
interpolate coarse grid data in time to provide boundary
conditions for fine grids.  Similarly, time interpolation may also be
required for interpolation or reduction operations at times when no
coarse grid data exist.  Such time interpolation is currently
implemented with up to fourth order accuracy, although only second
order is commonly used.

In order to achieve convergence at mesh refinement boundaries when
second spatial derivatives appear in the time evolution equations, we
do not apply boundary conditions during the substeps of a Runge--Kutta
time integrator.  Instead we extend the refined region by a certain
number of grid points before each time step (called \emph{buffer
  zones}) and only interpolate after each complete time step.  We find
that this is not necessary when no second derivatives are present, such as
in discretisations of the Euler equations without the Einstein
equations.
The details of this algorithm
are laid out in~\cite{Schnetter-etal-03b}.

Carpet is currently mostly used in situations where compact objects
need to be resolved in a large domain, and Carpet's mesh refinement
adaptivity is tailored for these applications.  One can define
several \emph{centres of interest}, and space will be refined around
these.  This is ideal e.g.\ for simulating binary systems, but is not
well suited for resolving shock fronts as they e.g.\ appear in
collapse scenarios.  We plan to address this soon in future versions
of Carpet.

\subsection{I/O}

Our I/O methods are based on the HDF5 \cite{hdf5web} library, which
provides a platform-independent high-performance file parallel format. 
Datasets are annotated with Cactus-specific descriptive attributes
 providing meta-data e.g.\ for the coordinate systems, tensor
types, or the mesh refinement hierarchy.

Some file systems can only achieve high performance when
each processor or node writes its own file. However, writing one-file-per-processor
on massively concurrent systems will be bottlenecked by the metadata server,
which often limits file-creation rates to only hundreds of files per second. 
Furthermore, creation of many thousands of files can create considerable
metadata management headaches. Therefore the most efficient
output setup is typically to designate every $n$th processor as output
processor, which collects data from $n-1$ other processors and writes
the data into a file.

Input performance is also important when restarting from a checkpoint.
We currently use an algorithm where each processor reads data only for
itself, trying to minimise the number of input file which it examines.

Cactus also supports novel server-directed I/O methods such as PANDA,
and other I/O file formats such as SDF and FlexIO.
All of these I/O methods are interchangeable modules that can be 
selected by the user at compile-time or runtime depending on their
needs and local performance characteristics of the cluster file systems.
This makes it very easy to provide head-to-head comparisons between
different I/O subsystem implementations that are nominally writing out
exactly the same data.

\section{Scaling on current machines}

We have evaluated the performance of the kernels of our applications on
various contemporary machines to establish the current state of the
code.  This is part of an ongoing effort to continually adapt the code to new
architectures.  These benchmark kernels include the time evolution
methods of
\codename{Ccatie} with and without \codename{Whisky}, using as driver
either \codename{PUGH} or \codename{Carpet}.
We present some recent benchmarking result below; we list the benchmarks
in table \ref{tab:benchmarks} and the machines and their
characteristics in table \ref{tab:machines}.

\begin{table}
  \caption{Our benchmarks and their characteristics.  The
    \codename{PUGH} and \codename{Carpet\_1lev} benchmarks evolve the
    vacuum Einstein equations without
    mesh refinement, with identical setups but using different
    communication strategies.
    \codename{Carpet\_8lev} features 8 fixed levels
    with the same number of grid points on each level.
    \codename{Whisky\_8lev} evolves the relativistic Euler equations
    in addition to the Einstein equations.
    \codename{BenchIO\_HDF5\_80l} writes several large files to disk
    using the Cactus checkpointing facility.}
  \begin{tabular}{l|lll}\hline
    Name                          & type    & complexity & physics \\\hline
    Bench\_Ccatie\_PUGH           & compute & unigrid    & vacuum  \\
    Bench\_Ccatie\_Carpet\_1lev   & compute & unigrid    & vacuum  \\
    Bench\_Ccatie\_Carpet\_8lev   & compute & AMR        & vacuum  \\
    Bench\_Ccatie\_Whisky\_Carpet & compute & AMR        & hydro   \\
    BenchIO\_HDF5\_80l            & I/O     & unigrid    & vacuum  \\\hline
  \end{tabular}
  \label{tab:benchmarks}
\end{table}

\begin{table}
  \caption{The machines used for the benchmarks in this section.  Note
    that we speak of ``processes'' as defined in the MPI standard;
    these are implicitly mapped onto the hardware ``cores'' or
    ``CPUs''.}

  %
  \begin{tabular}{ll|lll|rrrrr}\hline
    Name    & Host & CPU        & ISA    & Interconnect & \# proc  & cores/ & cores/ & memory/ & CPU freq. \\      
            &      &            &        &              &          &   node & socket &    proc &           \\\hline
    Abe     & NCSA & Clovertown & x86-64 & InfiniBand   &     9600 &      8 &      4 & 1 GByte &  2.30 GHz \\      
    Damiana & AEI  & Woodcrest  & x86-64 & InfiniBand   &      672 &      4 &      2 & 2 GByte &  3.00 GHz \\      
    Eric    & LONI & Woodcrest  & x86-64 & InfiniBand   &      512 &      4 &      2 & 1 GByte &  2.33 GHz \\      
    Pelican & LSU  & Power5+    & PPC64  & Federation   &      128 &     16 &      2 & 2 GByte &  1.90 GHz \\      
    Peyote  & AEI  & Xeon       & x86-32 & GigaBit      &      256 &      2 &      1 & 1 GByte &  2.80 GHz \\\hline
  \end{tabular}
  \label{tab:machines}
\end{table}

Since the performance of the computational kernel does not depend on
the data which are evolved, we choose trivial initial data for our
spacetime simulations, namely the Minkowski spacetime (i.e., vacuum).
We perform no analysis on the results and perform no output.  We
choose our resolution such that approximately 800~MByte of main memory
is used per process, since we presume that this makes efficient use of
a node's memory without leading to swapping.
We run the simulation for several time steps
requiring several wall-time minutes.
We increase the number of grid points with the number of processes.
We note that the typical usage
model for this code favors weak scaling as resolution of the computational
mesh is a major limiting factor for accurate modeling of the most demanding
problems.

Our benchmark source code, configurations, parameter files, and
detailed benchmarking instruction are available from the Cactus web
site
\cite{cactusweb1}.\footnote{See
  \url{http://www.cactuscode.org/Benchmarks/}} There we also present
results for other benchmarks and machines.

\subsection{Floating point performance}

\begin{figure}
  \includegraphics[width=0.45\textwidth]{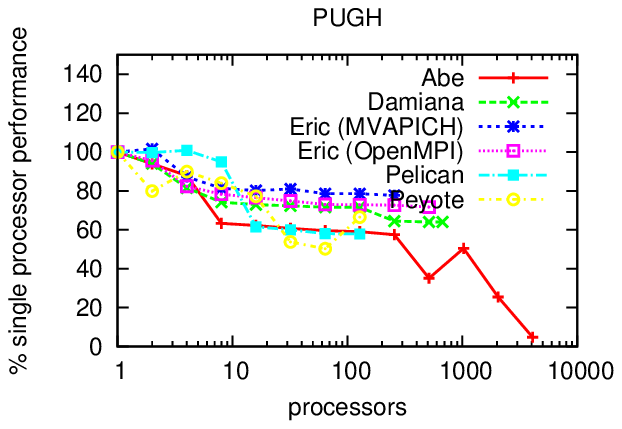}
  \includegraphics[width=0.45\textwidth]{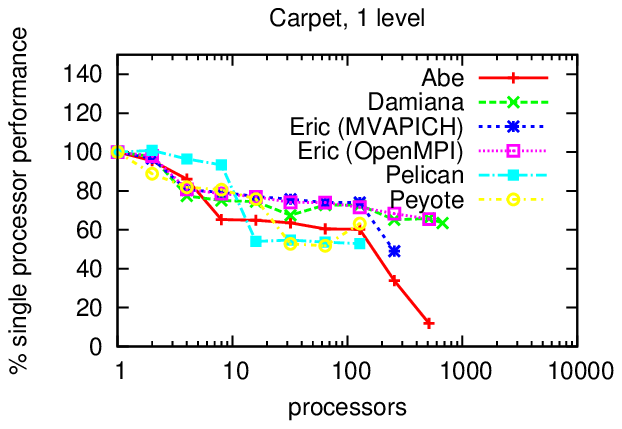}

  \includegraphics[width=0.45\textwidth]{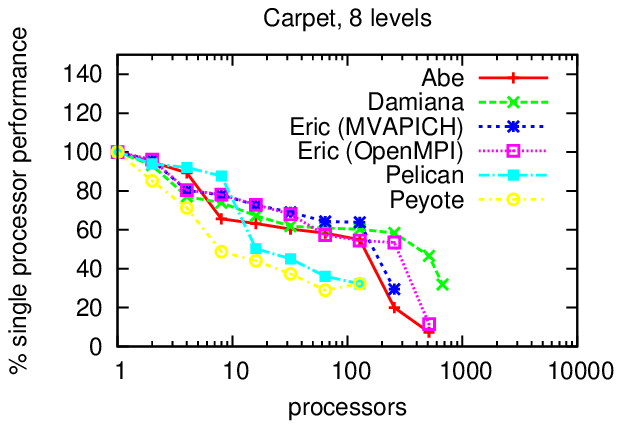}
  \includegraphics[width=0.45\textwidth]{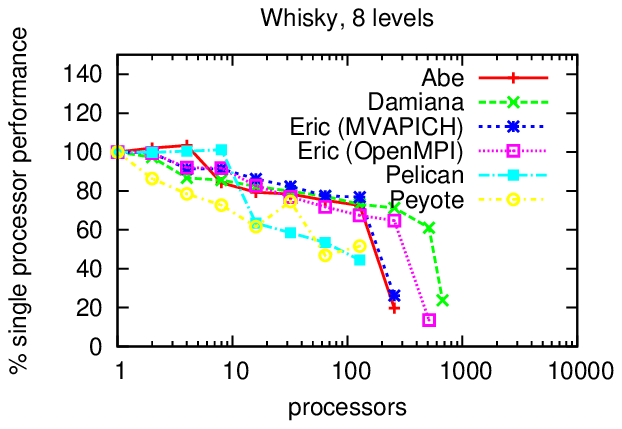}

  \caption{Weak scaling tests using our four benchmarks on various
    machines.  The graphs show the fraction of single processor
    performance which is achieved on multiple processors.
    The \codename{PUGH} and \codename {1lev} benchmarks
    have identical setups, but use different drivers with different
    communication strategies.  On Eric, we also compare two different
    MPI implementations.}
  \label{fig:scaling-benchmarks-weak}
\end{figure}

Figure \ref{fig:scaling-benchmarks-weak} compares the scaling
performance of our code on different machines.  The graphs show
scaling results calculated from the wall time for the whole simulation,
but excluding startup and shutdown.  The ideal result in all graphs is a
straight horizontal line, and larger numbers are better.  Values near
zero indicate lack of scaling.

Since the benchmarks \codename{Bench\_Ccatie\_PUGH} and
\codename{Bench\_Ccatie\_Carpet\_1lev} use identical setups, they
should ideally also exhibit the same scaling behaviour.
However, as the graphs
show, PUGH scales e.g.\ up to 1024 processors on Abe, while Carpet
scales only up to 128 processors on the same machine.  The differences
are caused by the different communication strategies used by PUGH and
Carpet, and likely also by the different internal bookkeeping
mechanisms.  Since Carpet is a mesh refinement driver, there is some
additional internal overhead (not communication overhead), even in unigrid simulations.

Both PUGH and Carpet use \codename{MPI\_Irecv}
\codename{MPI\_Isend}, and \codename{MPI\_Wait} to exchange data asynchronously.
PUGH exchanges ghost zone information in 3 sequential steps, first in
the $x$, then in the $y$, and then in the $z$ direction.  In each
step, each processor communicates with two of its neighbours.  Carpet
exchanges ghost zone information in a single step, where each processor
has to communicate with all 26 neighbours.
In principle, Carpet's algorithm should be more
efficient, since it does not contain several sequential steps;
in practice, performance suffers because it sends are larger number of small
messages than PUGH. The overhead of sending small messages has a marked impact on performance.  A lighter-weight communication mechanism such as one-sided messages may provide significant benefits to future implementations of Carpet.  

The fall-off at small processor numbers coincides with the node
boundaries.  For example, Abe drops off at the 8-processor boundary, and
Pelican drops off at the 16-processor boundary.  This is most likely
because a fully loaded node provides less memory bandwidth to each
process.

The scaling drop-off at large processor numbers for the commodity clusters
appears to be caused by
dramatic growth in memory usage by MPI implementations as the concurrency
of the simulation is scaled up. 
This drop-off is different for different MPI
implementations, as is e.g.\ evident on Eric, where OpenMPI scales
further than MVAPICH\@.  As the memory footprint grows, there is additional
TLB thrashing -- particularly on the Opteron processors, which have a comparatively
small TLB coverage of 2MB when using small pages.  
So it may only take a 1\% miss rate to
have a dramatic effect on performance. Large pages seem to not offer
a performance advantage.  

However, we saw now such growth in memory consumption on the BlueGene/L
system, which was able to scale up to 32k processors with little degradation as
shown in figure~\ref{fig:pugh-scaling}.
The BlueGene/L scaling study shows that by using a MPI implementation that
does not grow with concurrency, Cactus can continue to scale up to phenomenal
concurrencies with little drop-off in performance.

\subsection{I/O performance}

\begin{figure}
  \includegraphics[width=0.45\textwidth]{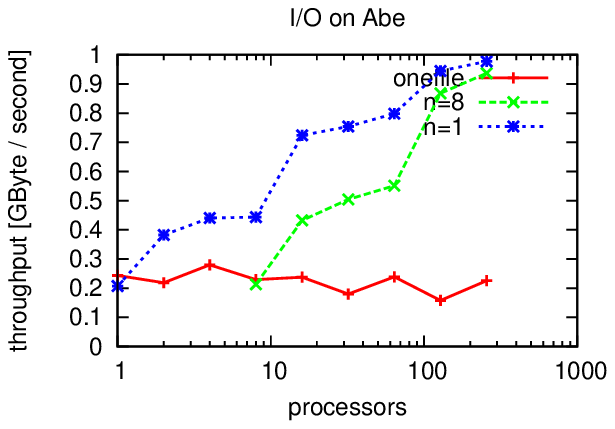}
  \includegraphics[width=0.45\textwidth]{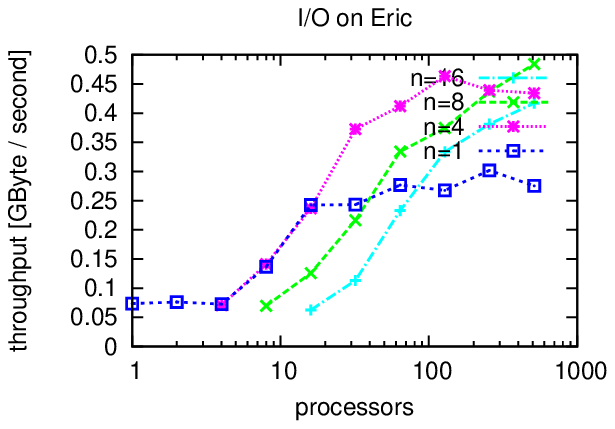}
  \caption{I/O scaling on various machines.  This benchmark was only
    run on Abe and Eric.  For each processor, 369~MByte of data are
    written to disk.}
  \label{fig:scaling-io}
\end{figure}

Figure \ref{fig:scaling-io} compares the scaling of different I/O
methods: Collecting the output onto a single processor, having every
$n$th processor perform output, and outputting on every processor.
Each output processor creates one file.  In these benchmarks, for each
compute processor 369~MByte need to be written to disk.  There is a
small additional overhead per generated file which is less than
1~MByte.

The throughput graphs of Abe and Eric show that the maximum
single-node throughput is limited, but the overall throughput can be
increased if multiple nodes write simultaneously.  While Abe has
sufficient bandwidth to support every CPU writing at the same time, it
is more efficient on Eric to collect the data first onto fewer
processors.

\section{Developing for petascale}

Petascale computing requires additional ingredients over conventional
parallelism.  The large number of available processors can only be
fully exploited with mechanisms for significantly more fine-grained
parallelism, in particular more fine grained than offered by the MPI
standard's collective communication calls.  In order to ensure
correctness in the presence of more complex codes and of increasingly
likely hardware failures, simulations should be divided into
\emph{parcels} describing a certain amount of computation.  These
parcels of computation can then be autonomously and dynamically distributed over the
available cores on the system by the Cactus \emph{driver}, and can be re-scheduled in response to hardware
failure, thus achieving both load balancing and end-to-end
correctness, akin to the way the IP protocol is used to deliver
reliability over unreliable components. Furthermore, the current
static simulation schedule --- often described by a sequence of
subroutine calls --- needs to become dynamic.  The latter can be
achieved e.g.\ by listing pre- and postconditions of each substep, and
determining an efficient execution order only at run time, akin to the
way UNIX \codename{make} functions.  The fact that Cactus physics modules
are written with the same rules as operators in functional languages 
(namely that they are stateless and have no side-effects), enables
this kind of flexibility in scheduling that will be necessary to carry us
beyond the Petascale systems. 

Clearly, these are only some measures, and a reliable methodology for
petascale computing requires a comprehensive set of concepts which
enable and encourage the corresponding algorithms and programming
styles.  It will very likely be necessary to develop new algorithms
for petascale machines.  These algorithms may not necessarily
run efficiently on single processors or today's popular small
workstation clusters, in the same way in which vectorised codes do not
work well on non-vector machines, or in which MPI codes may not work
well on single-processor machines.

One example of a non-petascale parallelisation algorithm which is
currently widely employed is the use of ghost zones to distribute
arrays onto multiple processors.  As the number of processors
increases, each processor receives a correspondingly smaller part of
the original array, while the \emph{overhead} (the ratio between the
number of ghost zones to the number of owned points) increases.  Some
typical numbers illustrate this: Our fourth-order stencils require a
ghost zone layer which is $3$ points wide.
If a processor owns $20^3$ grid points of a 3D array,
the overhead is $(26^3-20^3)/20^3 \approx 20\%$.  If a processor owns
only $10^3$ grid points, then the overhead increases to
$(16^3-10^3)/10^3 \approx 310\%$: the number of ghost zones is more
than three times the number of owned points.  Clearly we cannot
continue to scale this approach in its current form.  

Another problem for petascale simulations is time integration, i.e.,
the solution of hyperbolic equations.  Since each time step depends on
the result of the previous time step, time integration is inherently
serial.  If a million time steps are required, the total time to
solution is at least a million times the time for a single time step.

The Cactus framework with its clear division between physical
calculations and computational infrastructure already provides a
partial pathway to petascale computations.  By improving or exchanging
the driver, existing physics modules can be used in novel ways, e.g.\
invoking them on smaller parts of the domain, or scheduling several at
the same time to overlap computation and communication.  Such
improvement would be very difficult without a computational framework.
However, achieving full petascale performance will also require
improvements to the framework itself, i.e., to the way in which a
physics calculation describes itself and its interface to the
framework.

\subsection{Physics: radiation transport}

The most conceptually and technically challenging part of the
petascale development will be related to the implementation of neutrino
and photon radiation transport. While photons play an important role
in the jet propagation and, naturally, in the gamma-ray emission
of the GRB, neutrinos are of paramount importance in the 
genesis and evolution of the GRB engine, in particular in the
collapsar context. During the initial gravitational collapse to a 
protoneutron star and in the latter's short cooling period
before black hole formation, neutrinos carry away $\sim$99\% of
the liberated gravitational binding energy. After black-hole
formation, neutrino-cooling of the accretion disk and 
polar neutrino-antineutrino pair-annihilation are likely to be
key ingredients for the GRB mechanism. 

Ideally, the neutrino
radiation field should be described and evolved via the
Boltzmann equation, making the transport problem 7-dimensional
and requiring the inversion of a gigantic semi-sparse
matrix at each time step.
This matrix inversion
will be at the heart of the difficulties associated with
the parallel scaling of radiation transport algorithms and 
it is likely that even highly-integrated low-latency petascale
systems will not suffice for full Boltzmann radiation
transport, and sensible approximations will have to be worked
out and implemented.

One such approximation may be
neutrino, multi-energy-group,flux-limited diffusion (MGFLD)
along individual radial rays that are not coupled with each
other and whose ensemble covers the entire sphere/grid with
reasonable resolution of a few degrees. When in addition
energy-bin coupling (downscattering of neutrinos, 
a less-than-10\%-effect) and neutrino flavor changes are 
neglected, each ray for each energy group can be 
treated as a mono-energetic spherically symmetric calculation.
Each of these rays can then be domain decomposed, and the
entire ensemble of rays and energy groups can be updated
in massively-parallel and entirely scalable petascale fashion.

A clear downside of MGFLD is that all local angular
dependence of the radiation field is neglected, making it
(for example) fundamentally difficult to consistently estimate
the energy deposited by radiation-momentum angle-dependent
neutrino-antineutrino annihilation.
An alternative to the MGFLD approximation that can
provide an exact solution to the Boltzmann transport
equation, while maintaining scalability, is the statistical
Monte Carlo approach that follows a significant set of sample
particle random walks. Monte Carlo may also be the method
of choice for photon transport.

\subsection{Scalability}

In its current incarnation, the Cactus framework and its drivers PUGH
and Carpet are not yet fully ready for petascale applications.  Typical
petascale machines will have many more processing units than today's
machines, with more processing units per node, but likely with less
memory per processing unit.

One immediate consequence is that it will be impossible to
replicate metadata across all processors.  Not only the simulation
data themselves, but also all secondary data structures
will need to be distributed,
and will have to allow asynchronous remote access.  The driver thorns
will have to be adapted to function without global knowledge of the
simulation state, communicating only between neighbouring processors,
making global load balancing a challenge.

The larger number of cores per node will make hybrid communication
schemes feasible and necessary, where intra-node communication uses
shared memory while inter-node communication remains message based.
Explicit or implicit multithreading within a node will reduce the
memory requirements, since data structures are kept only once per
node, but will require a more complex overall orchestration to keep
all threads well fed.
Multithreading will have direct consequences for and
may require alterations to almost all existing components.
Multithreading will required a change of programming paradigm, as
programmers will have to avoid race conditions, and debugging will be
much more difficult than for a single-threaded code.

With the increasing number of nodes, globally synchronous operations
will become prohibitively expensive, and thus the notion of ``the
current simulation time'' will need to be abolished.  Instead of
explicitly stepping a simulation through time, different parts will
advance at different speeds.  Currently global operations, such as
reduction or interpolation operations, will need to be broken up into
hierarchical pieces which are potentially executed at different times, where the
framework has to ensure that these operations find the corresponding
data, and that result of such operations are collected back where they
were requested.

Current Cactus thorns often combine the physics equations which are to
be solved and the discretisation methods used to implement them, e.g.\
finite differences on block-structured grids.  Petascale computing may
require different discretisation techniques, and it will thus be
necessary to isolate equations from their discretisation so that one
can be changed while keeping the other.  Automated code generation
tools such as e.g.\ Kranc \cite{Husa:2004ip, krancweb} can
automatically generate physics modules from given equations, achieving
this independence. 

As a framework, Cactus will not implement solutions to these problems
directly, but will instead provide abstractions which can be implemented
by a variety of codes.  As external AMR drivers mature and prove
themselves, they can be connected into Cactus.  We are currently
engaged in projects to incorporate the PARAMESH \cite{parameshweb,
  MacNeice00} and SAMRAI \cite{samraiweb} AMR infrastructures
into Cactus drivers in the projects \emph{Parca} (funded by NASA)
and \emph{Taka}, respectively.

Instead of continuing to base drivers on MPI, we plan to investigate
existing new languages such as e.g.\ co-array Fortran, Titanium, and
UPC to provide parallelism, while we will also examine novel parallel
computing paradigms such as ParalleX\footnote{See
  \url{http://www.cs.sandia.gov/CSRI/Workshops/2006/HPC_WPL_workshop/Presentations/22-Sterling-ParalleX.pdf}}.
Cactus
will be able to abstract most of the differences between these, so
that the same physics and/or numerics components can be used with
different drivers.

It should be noted that the Cactus framework does not prescribe a
particular computing model.  For example, after introducing AMR
capabilities into Cactus, most existing unigrid thorns could be
AMRified with relatively little work; the existing abstraction that
each routine works on a block-structured set of grid point was
sufficient to enable the introduction of mesh refinement.  We expect
that the existing abstractions will need to be updated for petascale
computing, containing e.g.\ information about simultaneous action of
different threads and removing the notion of a global simulation time,
but we also expect that this same abstraction will then cover a
multitude of possible petascale computing models.

\subsection{Tools}

Petascale computing provides not only new challenges for programming
methodologies, but will also require new tools to enable programmers
to cope with the new hardware and software infrastructure.  Debugging
on $100,000+$ processors will be a formidable challenge, and only good
and meaningful profiling information for the new dynamic algorithms
will make petascale performance possible. These issues are being addressed
in three NSF-funded projects.

In the \emph{ALPACA} (Application Level Profiling And Correctness
Analysis) project, we are  developing interactive tools to debug and
profile scientific applications at the petascale level.  When a
code has left the stage where it has segmentation faults, it is
still very far from giving correct physical answers.  We plan to
provide debuggers and profilers which will not examine the programme
from the outside, but will run together with
the programme, being coupled to the programme via the framework, so
that it has first-class information about the current state of the
simulation.  We envision a smooth transition between debugging and
watching a production run progressing, or between running a special
benchmark and collecting profiling information from production runs on
the side.

The high cost of petascale simulations will make it necessary to treat
simulation results similar to data gathered in expensive experiments.
Our \emph{XiRel} (CyberInfrastructure for Numerical Relativity)
project seeks to provide scalable adaptive mesh refinement on one
hand, but also to provide means to describe and annotate simulation
results, so that
these results can later be interpreted and analysed unambiguously and
that the provenance of these data remains clear.  With ever increasing
hardware and software complexity, ensuring reproducibility of
numerical results is becoming an important part of scientific
integrity.

The \emph{DynaCode} project is providing capabilities and tools to
adapt and respond to changes in the computational environment, such as
e.g.\ hardware failures, changes in the memory requirement of the
simulation, or user-generated requests for additional analysis or
visualisation.  This will include
interoperability with other frameworks.

\section*{Acknowledgements}

We acknowledge the many contributions of our colleagues in the Cactus,
Carpet, and Whisky groups, and the research groups at the CCT, 
AEI, and at the University of Arizona.
We thank especially Luca Baiotti and Ian Hawke for sharing their
expertise on the Whisky code, and Maciej Brodowicz and Denis Pollney
for their help with the preparation of this manuscript.
This work has been supported by the Center for Computation \&
Technology at Louisiana State University, the Max-Planck-Gesellschaft,
EU Network HPRN-CT-2000-00137, NASA CAN NCCS5-153, NSF PHY 9979985, NSF 0540374
and by the Joint Institute for Nuclear
Astrophysics sub-award no.~61-5292UA of NFS award no.~86-6004791.

We acknowledge the use of compute resources and help from the
system administrators at the
AEI (Damiana, Peyote), LONI (Eric), LSU (Pelican), and NCSA (Abe).

\bibliographystyle{bibtex/apsrev}
\bibliography{bibtex/references,local_references}

\end{document}